\documentclass[11pt,a4paper,twoside,openright]{article}
\begin{document} 

\title{The Neutrino Oscillation Amplitudes\\[8pt] \Large Lecture Note}
\author{Walter Schmidt-Parzefall \\ Institut f\"ur Experimentalphysik, Universit\"at Hamburg}
\maketitle
\begin{abstract}
A consistent description of the results of the neutrino oscillation experiments carried out so far can be obtained from three basic properties of neutrinos and the parameters of the Hamiltonian matrix. Using these basic neutrino properties, an exact relation for the oscillation amplitudes of moving neutrinos is derived.
\end{abstract}

After the discovery of neutrino oscillations [1] it became clear, that the well known three neutrinos are not independent elementary particles, but transform into each other [2]. So far, for the analysis of neutrino oscillation experiments only approximate relations [3] are available. By the present note an exact relation for the oscillation amplitudes of moving neutrinos is derived from three basic properties of neutrinos and the parameters of the Hamiltonian matrix. The basic properties are:\\

Property 1

The three experimentally observable neutrino states $|\nu_e \rangle$, $|\nu_\mu \rangle$ and $|\nu_\tau \rangle$ are called flavor eigenstates. A neutrino $|\mbox{\boldmath$\nu$}\rangle$ is represented in terms of the flavor eigenstates $|\nu_n \rangle$ by
\begin{equation} |\mbox{\boldmath$\nu$}\rangle = \left( \begin{array}{c} |\nu_1 \rangle \\ |\nu_2 \rangle \\ |\nu_3 \rangle \end{array} \right) = \left( \begin{array}{c} |\nu_e \rangle \\ |\nu_\mu \rangle \\ |\nu_\tau \rangle \end{array} \right) \, . \end{equation}
A neutrino is a three-dimensional object in an internal flavor space.\\
 
Property 2

In general a flavor eigenstate depends on space and time coordinates, but there exists a special Lorentz frame, in which the flavor eigenstates only depend on the time $\tau$. This Lorentz frame is the CM system of a neutrino.\\

According to quantum mechanics, in the CM system the time evolution of the flavor eigenstates is given by a unitary transformation
\begin{equation} |\mbox{\boldmath$\nu$} (\tau) \rangle = \mbox{e}^{-\mbox{\footnotesize i} H (\tau - \tau_0)} |\mbox{\boldmath$\nu$} (\tau_0) \rangle \, . \end{equation}
As the generator of a unitary transformation, the Hamiltonian matrix $H$ is Hermitian and it is time-independent. Thus the time evolution of a neutrino is in the CM system described by the Schr\"odinger equation
\begin{equation} \mbox{i}\, \frac{\partial}{\partial \tau} \, |\mbox{\boldmath$\nu$} \rangle = H |\mbox{\boldmath$\nu$}\rangle \, . \end{equation}
As a Hermitian matrix, $H$ is diagonalized by a unitary matrix $U$
\begin{equation} H = U^{-1} M U \, , \end{equation} 
where $M$ is the mass matrix containing the mass eigenvalues $m_k$ 
\begin{equation} M = \left( \begin{array}{ccc} m_1 & 0 & 0 \\ 0 & m_2 & 0 \\ 0 & 0 & m_3 \end{array} \right) . \end{equation}
The matrix $U$ depends on four independent parameters, three rotation angles and one complex phase, and the matrix $M$ depends on three independent parameters, the mass eigenvalues $m_k$. The Hamiltonian matrix $H$ is completely described by these parameters. Within the framework of the Standard Model of elementary particle physics the Hamiltonian matrix cannot be derived. It is the task of neutrino experiments to determine its parameters.\\

The solutions of the Schr\"odinger equation can be obtained [2] by the transformation
\begin{equation} |\mbox{\boldmath$\nu$}^\prime \rangle = U |\mbox{\boldmath$\nu$} \rangle   \, , \end{equation}
where the mass eigenstates $|\nu_k^\prime \rangle$ are the components of $|\mbox{\boldmath$\nu$}^\prime \rangle$, and the mass eigenstates $|\nu_k^\prime \rangle$ and the flavor eigenstates $|\nu_n \rangle$ are transformed by
\begin{equation} |\nu_k^\prime \rangle = \sum_n U_{kn} |\nu_n \rangle   \qquad \mbox{and} \qquad |\nu_n \rangle = \sum_k  U^{-1}_{\quad nk} |\nu_k^\prime \rangle = \sum_k U^*_{kn} |\nu_k^\prime \rangle \, . \end{equation}
With (4) and (6) the Schr\"odinger equation turns into the eigenvalue equation
\begin{equation} \mbox{i}\, \frac{\partial}{\partial \tau} \, |\mbox{\boldmath$\nu$}^\prime \rangle = M |\mbox{\boldmath$\nu$}^\prime \rangle \, , \qquad \mbox{where} \qquad \mbox{i}\, \frac{\partial}{\partial \tau} \, |\nu_k^\prime \rangle = m_k |\nu_k^\prime \rangle \, , \end{equation}
and the mass eigenstates are the solutions 
\begin{equation}|\nu_k^\prime (\tau) \rangle = \mbox{e}^{-\mbox{\footnotesize i} m_k (\tau - \tau_0)} |\nu_k^\prime (\tau_0) \rangle \, . \end{equation}

Property 3 

At the moment of creation or annihilation a neutrino takes on a pure flavor eigenstate.\\ 

The initial state $|\mbox{\boldmath$\nu$}_i (\tau_0) \rangle$ at time $\tau_0$ is a pure flavor eigenstate carrying the initial flavor $i = e, \mu,\tau$. Its components are the flavor eigenstates
\begin{equation} |\nu_{ni} (\tau_0) \rangle = \delta_{ni} |\nu_0 \rangle \, , \end{equation}
where $|\nu_0 \rangle$ is a neutrino state normalized to $\langle \nu_0 |\nu_0 \rangle = 1$.\\

Using (7) the initial flavor eigenstates $|\nu_{ni} (\tau_0) \rangle $ transform into the initial mass eigenstates
\begin{equation}|\nu_{ki}^\prime (\tau_0) \rangle = \sum_n U_{kn} |\nu_{ni} (\tau_0) \rangle = \sum_n U_{kn} \delta_{ni} |\nu_0 \rangle = U_{ki} |\nu_0 \rangle \, , \end{equation}
and the time evolution of the mass eigenstates (9) results in
\begin{equation}|\nu_{ki}^\prime (\tau) \rangle = \mbox{e}^{-\mbox{\footnotesize i} m_k (\tau - \tau_0)} U_{ki} |\nu_0 \rangle \, . \end{equation}
Hence the time evolution of the flavor eigenstates is given by
\begin{equation} |\nu_{ni} (\tau) \rangle = \sum_k U^*_{kn} |\nu_{ki}^\prime (\tau) \rangle = \sum_k U^*_{kn} U_{ki} \mbox{e}^{-\mbox{\footnotesize i} m_k (\tau - \tau_0)}|\nu_0 \rangle \, . \end{equation}
Thus the probability amplitude $\mathcal{A}_{fi} (\tau) $ for the neutrino $|\mbox{\boldmath$\nu$}_i (\tau) \rangle$, produced with the initial flavor $i$, to be observed in the pure flavor eigenstate $|\mbox{\boldmath$\nu$}_f \rangle$ with the final flavor $f$ is obtained
\begin{equation}  \mathcal{A}_{fi} (\tau) = \langle \mbox{\boldmath$\nu$}_f |\mbox{\boldmath$\nu$}_i (\tau) \rangle = \sum_n \langle \nu_0 \delta_{fn} |\nu_{ni} (\tau) \rangle = \sum_k U_{kf}^* U_{ki} \mbox{e}^{-\mbox{\footnotesize i} m_k (\tau - \tau_0)} \, . \end{equation}
This well known expression [3] contains oscillation terms and describes neutrino oscillations in the CM system. Now we investigate neutrino oscillations of moving neutrinos.\\ 

Describing a free neutrino moving in $x$-direction by solutions of the Dirac equation, it is formed by plane waves, which can be represented by mass eigenstates
\begin{equation} |\nu^\prime_{ki} (x,t) \rangle = \mbox{e}^{-\mbox{\footnotesize i} \left( E_k (t - t_0 ) - p_k (x - x_0 )\right )} U_{ki} |\nu_0 \rangle \, , \end{equation}
where $E_k$ and $p_k$ are the energy and the momentum of a wave and $t$ and $x$ denote the coordinates in the Lab system. Requiring that these waves remain invariant, when the neutrino is described at rest by (12) or in motion by (15), implies
\begin{equation} m_k (\tau - \tau_0 ) = E_k (t - t_0 )- p_k (x - x_0 ) \, . \end{equation}
With $\gamma^2 - \beta^2 \gamma^2 = 1$, this requirement is fulfilled by the basic relations
\begin{equation} t - t_0 = \gamma (\tau - \tau_0 ) \, , \qquad x -x_0 = \beta \gamma (\tau -\tau_0 ) \end{equation}
and
\begin{equation} E_k = \gamma m_k \, , \qquad \qquad p_k = \beta \gamma m_k \, . \end{equation}

An oscillating neutrino is not an eigenstate of energy or momentum. However, it has a well defined energy expectation value $\langle E \rangle$ and momentum expectation value $\langle p \rangle$. In order to obtain the oscillation amplitude $\mathcal{A}_{fi} $ for moving neutrinos, the neutrino momentum expectation value $\langle p \rangle$ can be used. It is defined by
\begin{eqnarray} \langle p \rangle &=& \langle \mbox{\boldmath$\nu$}_i (x,t) | -  \mbox{i} \,\frac{\partial}{\partial x} | \mbox{\boldmath$\nu$}_i (x,t) \rangle  = \langle \mbox{\boldmath$\nu$}^\prime_i (x,t) \, U | -  \mbox{i} \,\frac{\partial}{\partial x} | U^{-1} \mbox{\boldmath$\nu$}^\prime_i (x,t) \rangle \nonumber\\
&=& \langle \mbox{\boldmath$\nu$}^\prime_i (x,t) | -  \mbox{i} \,\frac{\partial}{\partial x} | \mbox{\boldmath$\nu$}^\prime_i (x,t) \rangle \, , \end{eqnarray}
since $U$ does not depend on $x$. Using (15) and (18) results in
\begin{equation} \langle p \rangle = \sum_k \langle \nu^\prime_{ki} (x,t)| -  \mbox{i} \,\frac{\partial}{\partial x} | \nu^\prime_{ki} (x,t) \rangle = \sum_k U_{ki}^* \, p_k U_{ki} = \beta \gamma \sum_k |U_{ki}|^2 m_k \, . \end{equation}
Since due to (4) the diagonal elements of $H$ are
\begin{equation} H_{ii} = \sum_k |U_{ki}|^2 m_k \, , \end{equation}
we finally obtain
\begin{equation} \langle p \rangle = \beta \gamma H_{ii} \, . \end{equation}

The oscillation amplitude $\mathcal{A}_{fi} (\tau)$ which is given by equation (14) depends on the time difference $\tau - \tau_0$ elapsed in the CM system. For a moving neutrino this time difference can be expressed in terms of quantities observable in the Lab system. During $\tau - \tau_0$ a neutrino travelling with the velocity $\beta$ in the Lab system will move there over a distance $L$ given by (17)
\begin{equation} L = x - x_0 = \beta \gamma \, (\tau - \tau_0) \, .\end{equation}
The Lorentz factor $\beta \gamma$ can be obtained from (22) resulting in
\begin{equation} \tau - \tau_0 = H_{ii}\frac{L}{\langle p \rangle} \, .\end{equation}
Thus the oscillation amplitude $\mathcal{A}_{fi}$ can be rewritten in terms of quantities of the Lab system 
\begin{equation} \mathcal{A}_{fi} = \sum_k U_{kf}^* U_{ki} \, \mbox{e}^{-\mbox{\footnotesize i}\, m_k(\tau-\tau_0)} = \sum_k U_{kf}^* U_{ki} \, \mbox{e}^{-\mbox{\footnotesize i}\, m_k H_{ii} \textstyle \frac{L}{\langle p \rangle}}  \, . \end{equation}

This is the main result, derived from the three neutrino properties introduced before. It depends on seven arbitrary parameters, which have to be determined by experiment and leads in general to complicated expressions.\\

In many cases some simplifications can be made. As an example, an experimentally well justified approximation of $\mathcal{A}_{fi} $ for $\nu_\mu \leftrightarrow \nu_\tau$ oscillations is obtained by making two assumptions.\\

Assumption 1

With regard to neutrino oscillations, $\nu_\mu$ and $\nu_\tau$ are indistinguishable. The Schr\"odinger equation is invariant under exchange of a $\nu_\mu$ with a $\nu_\tau$. Thus $H$ is approximated by
\begin{equation} \hat{H} = \left( \begin{array}{ccc} A & B & B \\ B & C & D \\ B & D & C \end{array} \right) . \end{equation} 
The matrix $\hat{H}$ is diagonalized by 
\begin{equation} \hat{U} = \frac{1}{\sqrt{2}} \left( \begin{array}{rrc} \sqrt{2} \, c_1 & s_1 & s_1 \\ - \sqrt{2} \, s_1 & c_1 & c_1 \\ 0 \quad & -1 \, & 1 \end{array} \right) . \end{equation}
Choosing as a suitable parameterization of $U$
\begin{eqnarray} U &=& \left( \begin{array}{rcc} c_1 & s_1 & 0 \\ -s_1 & c_1 & 0 \\ 0 \: & 0 & 1 \end{array} \right) \left( \begin{array}{ccc} c_2 & 0 &  -s_2\mbox{e}^{-\mbox{\footnotesize i}\delta}  \\ 0 & 1 & 0 \\ s_2 \mbox{e}^{\mbox{\footnotesize i}\delta} & 0 & c_2 \end{array} \right) \left( \begin{array}{crc} 1 & 0 \: & 0 \\ 0 & c_3 & s_3 \\ 0 & -s_3 & c_3 \end{array} \right), \nonumber\\
& &s_n^2 + c_n^2 = 1, \end{eqnarray} 
the matrix $\hat{U}$ results from $U$ for the parameters
\begin{equation} s_2 = 0, \qquad s_3 = \frac{1}{\sqrt{2}} \, .\end{equation}
The experimental data [1], [4] obtained so far are consistent with these parameters.
Future experiments will show, to which extent this symmetry is realized.\\

Assumption 2

$\nu_\mu$ and $\nu_\tau$ form a two-state system. We switch off $\nu_e$ oscillations. Thus $B$ is approximated by 
\begin{equation} B = \frac{1}{\sqrt{2}} s_1 c_1 (m_1 - m_2) = 0 \, , \end{equation}
where (4) and (27) were used. Since we know from experiment [5], that $|m_2 - m_1 | \ll |m_3 - m_2 |$, we choose the parameters
\begin{equation} m_1 = m_2 \, ,\end{equation} 
whereas $s_1$ remains arbitrary, and obtain from (21) and (27)
\begin{equation} \hat{H}_{22} = \hat{H}_{33} = C = \frac{1}{2} ( s_1^2 m_1 + c_1^2 m_2 + m_3 ) = \frac{1}{2} ( m_2 + m_3 )  \, .\end{equation}
Inserting $\hat{U}$ and $m_1 = m_2$ into (25) results in the approximate oscillation amplitudes $\mathcal{A}_{fi}$ and the corresponding oscillation probabilities $P_{fi} = |\mathcal{A}_{fi} |^2$ for neutrinos moving over a distance $L$
\begin{eqnarray} 
\mathcal{A}_{\mu \tau} = \mathcal{A}_{\tau \mu} &=& \frac{1}{2} \left(s_1^2 \mbox{e}^{-\mbox{\footnotesize i}\, m_1 (\tau - \tau_0) } + c_1^2 \mbox{e}^{-\mbox{\footnotesize i}\, m_2 (\tau - \tau_0) } - \mbox{e}^{-\mbox{\footnotesize i}\, m_3 (\tau - \tau_0) } \right) \nonumber\\
&=& \frac{1}{2} \left( \mbox{e}^{-\mbox{\footnotesize i}\, m_2 (\tau - \tau_0) } - \mbox{e}^{-\mbox{\footnotesize i}\, m_3 (\tau - \tau_0) } \right) \nonumber\\
P_{\mu \tau} = P_{\tau \mu} &=& \sin^2 \left( \frac{m_2 - m_3}{2} (\tau - \tau_0) \right) = \sin^2 \left( \frac{m_2 - m_3}{2}\, \frac{m_2 + m_3}{2} \frac{L}{\langle p \rangle} \right) \nonumber\\ &=& \sin^2 \left( \frac{m_2^2 - m_3^2}{4}\, \frac{L}{\langle p \rangle} \right) \nonumber\\
\mathcal{A}_{\mu \mu} = \mathcal{A}_{\tau \tau} &=& \frac{1}{2} \left(s_1^2 \mbox{e}^{-\mbox{\footnotesize i}\, m_1 (\tau - \tau_0) } + c_1^2 \mbox{e}^{-\mbox{\footnotesize i}\, m_2 (\tau - \tau_0) } + \mbox{e}^{-\mbox{\footnotesize i}\, m_3 (\tau - \tau_0) } \right) \nonumber\\
P_{\mu \mu} = P_{\tau \tau} &=& \cos^2 \left( \frac{m_2^2 - m_3^2}{4}\, \frac{L}{\langle p \rangle} \right) .
\end{eqnarray}
This example shows that equation (25) results in the well known simple relations for $\nu_\mu \leftrightarrow \nu_\tau$ oscillations, after the appropriate approximations for the oscillation parameters have been made. However, the main advantage of equation (25) is to allow for exact calculations of the amplitudes of neutrino oscillations.\\ 

{\bf Acknowledgements} \newline
It is a pleasure to thank Prof.\,C.\,Hagner and her neutrino group for kind hospitality and T.\,Ferber for clarifying discussions.\\

{\bf References} \newline 
[1] R.\,Davis, Prog. Part. Nucl. Phys. {\bf 32} 13 (1994);\\ \hspace*{12pt} 
S.\,Fukuda et al. (Super Kamiokande) Phys. Rev. Lett. {\bf81} 1562 (1998); \\ \hspace*{12pt} Q.\,R.\,Ahmad et al. (SNO) Phys. Rev. Lett. {\bf 89} (2002). \newline
[2] B. Pontecorvo, Sov. Phys. JETP {\bf7} 172 (1958); \\ \hspace*{12pt} 
Z.\,Maki, M.\,Nakagawa and S. Sakata, Prog. Theor. Phys. {\bf 28} 870 (1962).  \newline
[3] B.\,Kayser, in Phys. Lett. B {\bf 667} 163 (2008) and references therein.\newline
[4] M.\,Apollonio et al. (CHOOZ) Eur. Phys. J. C {\bf 27} 331 (2003); \\ \hspace*{12pt}
R.\,Ospanov et al. (MINOS) Fermilab-Thesis-2008-04 (2008). \newline
[5] T.\,Araki et al. (KamLAND) Phys. Rev. Lett. {\bf 94} 081801 (2005).
 
\end{document}